\documentclass[conference]{IEEEtran}
\IEEEoverridecommandlockouts
\usepackage{cite}
\usepackage{amsmath,amssymb,amsfonts}
\usepackage{algorithmic}
\usepackage{graphicx}
\usepackage{textcomp}
\usepackage{xcolor}
\usepackage{float}
\usepackage{adjustbox}
\usepackage{booktabs}

\def\BibTeX{{\rm B\kern-.05em{\sc i\kern-.025em b}\kern-.08em
    T\kern-.1667em\lower.7ex\hbox{E}\kern-.125emX}}
\begin{document}

\title{Exploring Lossy Compressibility through Statistical Correlations of Scientific Datasets}

\author{\IEEEauthorblockN{David Krasowska\IEEEauthorrefmark{1}, Julie Bessac\IEEEauthorrefmark{2}, Robert Underwood\IEEEauthorrefmark{1}, Jon C. Calhoun\IEEEauthorrefmark{1}, Sheng Di\IEEEauthorrefmark{2}, and Franck Cappello\IEEEauthorrefmark{2}
\IEEEauthorblockA{\IEEEauthorrefmark{1} \textit{Holcombe Department of Electrical and Computing Engineering}, \textit{Clemson University} Clemson, USA}
\IEEEauthorblockA{\IEEEauthorrefmark{3} \textit{Mathematics and Computer Science Division}  \textit{Argonne National Laboratory} Lemont, USA}
\IEEEauthorblockA{\IEEEauthorrefmark{3} \textit{School of Computing}, \textit{Clemson University} Clemson, USA }}

}

\maketitle

\begin{abstract}
Lossy compression plays a growing role in scientific simulations where the cost of storing their output data can span terabytes. Using error bounded lossy compression reduces the amount of storage for each simulation; however, there is no known bound for the upper limit on lossy compressibility. Correlation structures in the data, choice of compressor and error bound are factors allowing larger compression ratios and improved quality metrics. Analyzing these three factors provides one direction towards quantifying lossy compressibility. As a first step,  we explore statistical methods to characterize the correlation structures present in the data and their relationships, through functional models, to compression ratios. We observed a relationship between compression ratios and statistics summarizing correlation structure of the data, which are a first step towards evaluating the theoretical limits of lossy compressibility used to eventually predict compression performance and adapt compressors to correlation structures present in the data. 
\end{abstract}


\section{Introduction}

Scientific research increasingly uses error-bounded lossy compressors to achieve greater compression ratios in relation to lossless compressors\cite{Cappello2019}. This improved performance allows applications to run with larger and more frequently produced datasets due to faster I/O times and smaller I/O volumes. The theoretical limit on compressibilty of data using lossless compression is given by the entropy\cite{Shannon1948}. The entropy quantifies the information content present in a symbol from a source sequence based upon its probability of occurrence. Thus, for a given sequence of symbols the entropy enables computing the minimum number of bits, on average, needed to represent the data. 
For over 70 years, this concept has guided the development and evaluation of lossless compression algorithms. However, for lossy compression algorithms, there is currently no known bound for the maximum degree of compression that can be achieved for some specified point-wise error bound. Establishing this bound will allow researchers to anticipate compression performance and adapt compressors to correlation structures of the data ensuring they get the best compression ratio possible.
Similarly, establishing the limit for lossy compression allows for the maximum efficiency for storing large scientific datasets. 

One possible direction to establish an \textit{entropy-like} bound for lossy compression is studying the impact on compressibility of correlation structures (correlations in space, time, or other dimensions) of the data, compressor types and error bounds.  
We refer to ``compressibility" as the maximum compression ratio associated with a given error bound. 
As many compressors implicitly or explicitly exploit correlation structures of the data (e.g.  SZ \cite{Liang2018}, 
ZFP \cite{Lindstrom2006}), analyzing the relationships between correlation structures and compression performances will allow researchers to anticipate compression performances. 
Ultimately, we seek to establish an entropy-like metric for lossy compression algorithms which can guide the lossy compression community to optimal development and usage by adapting compressors to correlation structures present in the  data. 

In this work, we focus on data correlations and their link to compression ratios for several compressors.  
The goal of the research paper is to explore:
\begin{enumerate}
    \item statistical methods to characterize the correlation structures of the data and
    \item their relationships, through functional models, to compression ratios.  
\end{enumerate} 
These models will form the first step into evaluating the theoretical limits of lossy compressibility used to eventually predict compression performance. 
In particular, we focus on estimated correlation ranges through variograms and its effect on compressibility. The variogram is commonly used in geostatistcis to estimate second-order dependence and more precisely how data are correlated with distance. 
Under stationary conditions, the variogram and covariance function have a direct correspondence. However, in practice the variogram can be estimated in more relaxed assumptions than the  covariance and thus is preferred by practitioners. 
To characterize compressibility, we use the compression ratio,  which is an important statistic within lossy compression due to its positive correlation with storing and processing as much data as possible.

 In this study, we focus on 2D-gridded datasets and through variogram analysis, perform a characterization of the  correlation between  grid-points along with a compression analysis (Section \ref{sec:stats}) with several compressors (Section \ref{sec:compressor}). 
 We consider two types  of data (Section \ref{sec:data}), synthetic datasets consisting of stochastic correlated Gaussian fields with  known correlation structures and another one consisting of simulations from a hydrodynamical model Miranda \cite{capello_zhao_di_tao_bessac_chen}. 
 
\section{Background}

\subsection{Compressors}\label{sec:compressablity:compressors}
SZ \cite{Liang2018}, ZFP \cite{Lindstrom2006}, and MGARD \cite{Ainsworth2019} are some of the leading error bounded lossy compressors.
In this section we explain from an algorithmic perspective how these compressors exploit correlations in data.

SZ scans through the data block by block: for 2D data the block size is $16 \times 16$. For each block, it makes a prediction of what the data in the block is using one of two methods: the Lorenzo predictor or the regression predictor.  The Lorenzo predictor uses the values of neighboring points to estimate the value at the current position.  The regression predictor fits a hyper-plane through the block, and uses the fitted hyper plane to interpolate the values within the blocks.  If these predictions are linearly quantized, and if the quantized values are sufficiently accurate according to the error bound, the quantized values are stored.  If the predictions are not accurate, the values are stored exactly.
Finally the entire sequence is passed first to a huffman encoding, and then to the Zstd lossless compressor to exploit patterns in the quantized sequence.
However, since the the predictor does not observe values outside of its block, it cannot exploit global correlation structures easily.

ZFP likewise uses local correlations, but uses a completely different compression principle based on near orthogonal transforms -- a similar approach used by JPEG image compression.
ZFP first partitions 2D data into blocks of size $4 \times 4$, then it converts each block of floating point data into a common fixed point representation, applies the near orthogonal transform, applies an embedded encoding that orders bits from most significant to least significant and then applies a truncation to archive a desired tolerance.
Again, because the blocks are compressed independently, the compressor cannot acquire a global knowledge of the data's correlations structures.

MGARD however is a newer compressor uses a different approach which can account for global correlation structures.
MGARD relies on the mathematical theory of multi-grid methods in order to compress the data.
MGARD operates by decomposing the data into multi-level coefficients which represent recursively defined sub-regions until the block is represented within the allowed tolerance.
These multi-level coefficients are then quantized and then compressed with either Zlib (in older versions) or Zstd (in the newest unreleased version).
Because these multi-level coefficients can represent regions of differing sizes and potentially the entire dataset, MGARD can capture multi-level effects in a way that SZ and ZFP cannot making it an important comparison for our paper.

\subsection{Variogram}
\begin{figure}
\centering
\includegraphics[width=\linewidth]{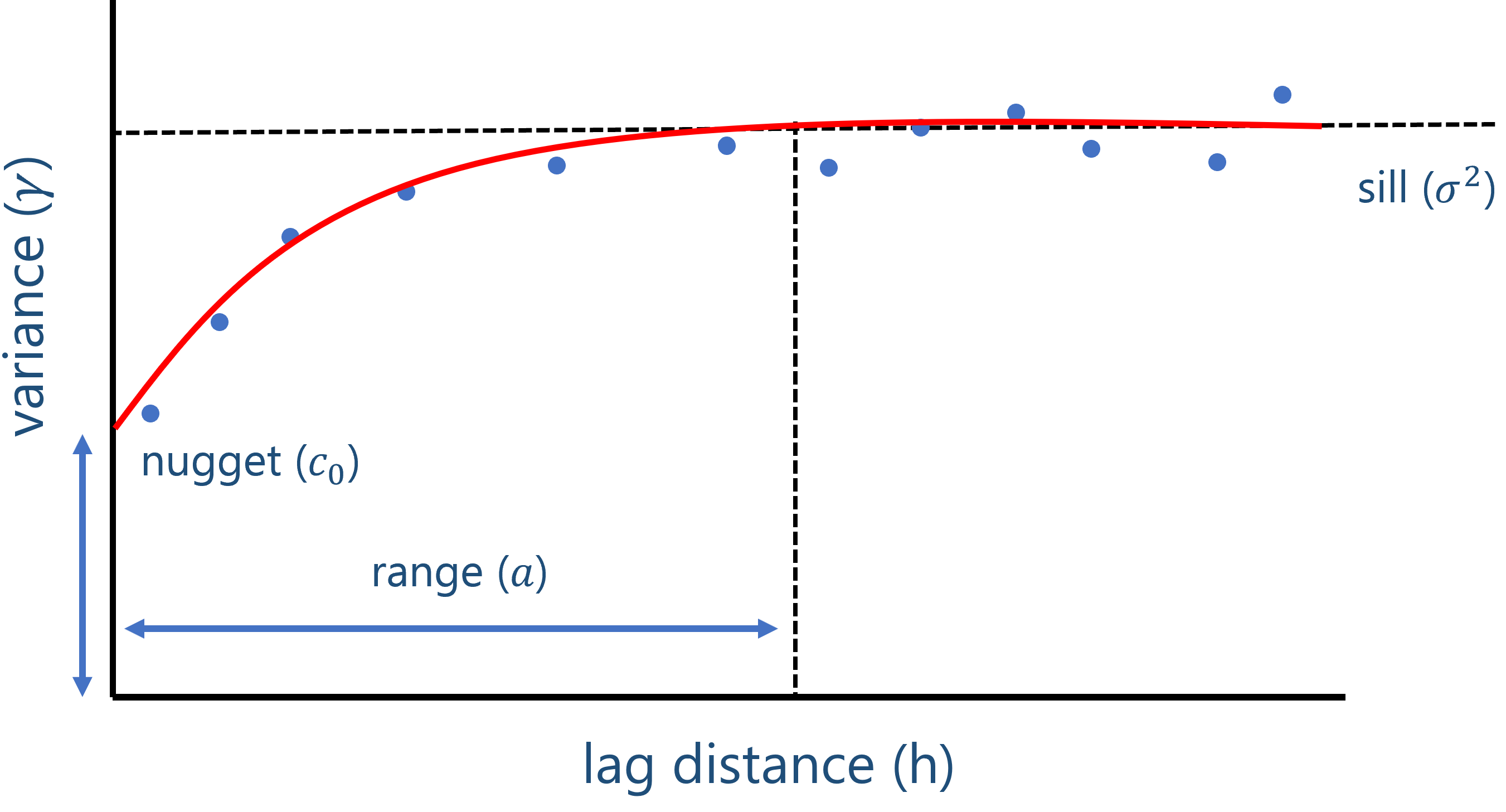}
\vspace{-20pt}
\caption{Example of a variogram as a function of the  distance $h$ between points. }
\label{fig:variogram}
\vspace{-10pt}
\end{figure}

The variogram is a function describing the degree of dependence of a correlated spatial field \cite{matheron1963, gelfand2010}, it gives a measure of how much two points of the same field correlate depending on the distance $h$ between those points. 
The variogram is a function of the distance $h$ and is typically characterized by several parameters: nugget  (microscale variability), sill (variance of the studied field), range (distance at which autocorrelation vanishes), see Figure \ref{fig:variogram}. 
%graphs_arxiv/
In the following, we focus solely on the variogram range $a$ as it corresponds to the distance ($h$) where the variogram ($\gamma$) plateaus (at the sill value) and indicates the distance beyond which the spatial correlation among grid-points vanishes.  
Intuitively, the larger the range the stronger the correlation is across grid-points. 

In practice, the empirical semi-variogram is computed on the data via Equation \eqref{eq:variogram}:
\begin{equation}\label{eq:variogram}
\displaystyle \gamma(h) = \frac{1}{2N(h)}\sum_{|x_i - x_j|=h}^{N(h)}\left ( z(x_i)-z(x_j) \right)^2,
\end{equation}
where $z$ is the studied field of interest (e.g. \textit{velocityx} from the Miranda dataset), $x_i$ and $x_j$ are grid-points coordinates or indexes, $N(h)$ is the number of grid-points at distance $h$ from each other. 
The variogram corresponds to $2\gamma(h)$, in practice the terms semi-variogram and variogram are used interchangeably. In the following, we will refer to variogram as $\gamma(h)$. 
Finally, to estimate the variogram range $a$ we fit by least-squares a parametric squared-exponential variogram ($\gamma(h) = c_{0} (1-\exp(-h^2/a)) $) to the empirical variogram range estimated via Equation \eqref{eq:variogram}.

\subsection{Complex correlation patterns} 
Statistical tools exist beyond the variogram to quantify and extract complex correlation structures of datasets for instance in order to analyze long-range dependencies \cite{abry1998}, to detect change-point in time  series \cite{cabrieto2017}, to cluster and identify correlation regimes \cite{bessac2016}. 
Identifying multiscale components of scientific datasets mostly relies on eigen or a basis-function decompositions  such  as  singular value decomposition (SVD) or wavelet decomposition \cite{addison2017,hannachi2007}. 
However due to the often extreme complexity of correlations and dependencies in data, developing methods to extract for instance spatial and spatio-temporal heterogeneity  or non-stationarity is still on-going research. 
Although the detailed use of these techniques is outside the scope of the current work and left for future efforts, we provide preliminary results  and future work  directions with SVD at the end of the Section~\ref{sec:compressability:results}.

\section{Related Work}
Beyond the classical efforts to estimate compressiblity using entropy, there has been relatively little attention afforded to the topic of lossy compressiblity.
As a parallel in data reduction techniques, \cite{gavish2014} investigated the determination of thresholds for singular value  decomposition of large matrices based on some optimality loss  criteria.   
\cite{lu2018understanding} identified several factors that affect compression ratio for SZ and ZFP.  
They relied on block-based sampling based approach which considered individual data points tailored to each considered compressor that they consider and used Gaussian model to estimate the subsequent compressibility. 
For SZ, they considered a number of input predictors such as the number of elements in the dataset, the quantization interval, information about the Huffman tree constructed, and the number of points that were unpredictable with SZ's predictor.
For ZFP, they offered a proof of their sampling methods estimated accuracy and empirically show it to be 99\% accurate across many datasets. 
In the same vain, \cite{qin2020estimating} based  on their prior work, used deep neural networks to estimate the compression ratio instead of a Gaussian model.
 However the built neural network may not generalize to other applications because it could simply over-learn the training data and the testing data would not prevent the over-learning \cite{hanDataMiningConcepts2011}. 
\cite{taoOptimizingLossyCompression2019} designed an automated methodology to switch between SZ and ZFP based on which compressor  is estimated to have a greater compression ratio. Compression ratios are estimated in a block-based sampling approach using Shannon entropy \cite{Shannon1948} of the sampled quantized blocks to investigate SZ's behavior. 
Most of these works have limited generalizability because of their reliance on algorithmic details of each studied compressor or reduction technique.

Finally, little effort has been directed to explore explicit links of correlation structures in the data to reduction and compression techniques and their  performance. 
\cite{lindstrom2014} has explored the decorrelation efficiency of a specific reduction method on scientific datasets in order to identify trade-offs for parameterization. 
\cite{moon2017} lead an evaluation and comparison of several lossy reduction techniques  based on  basis-function decompositions adapted  to temporal and spatial data. 
\cite{yu2021} developed an adaptive hierarchical geospatial field data representation
(Adaptive-HGFDR) based on blocked hierarchical
tensor decomposition to exploit multidimensional correlations on the data. However, none of these works systematically investigate the explicit link between correlation structures and compressibility. 
Our work goes beyond these approaches to consider a direction that is compressor independent by looking only at local spatial relationships in the data.
However, unlike these papers we leave to future work the task of estimating the compression ratio from the observations about spatial relations we study in this paper.

\section{Methodology}

\subsection{Datasets}\label{sec:data}
%All datasets are little endian IEEE 754 64 bit doubles. 
In the following, we refer to datasets as a particular field at a  particular time when in an application (e.g. a single continuous  variable  in memory). In particular,  we work with a single temporal snapshot of the studied data. 

The first dataset consists of synthetic 2D Gaussian fields with a controllable correlation structure following  a squared-exponential correlation model. These 2D fields are 1028$\times$1028 grid-points. 
We  consider these fields as ``ideal'' as the correlation range is known and varied to create multiple correlated fields. Gaussian fields $z$ over a grid defined by indexes $x_i$ are generated using the following probability distribution $f$
\begin{equation}\label{eq:pdf_gauss}
    f(z(x_{1}),\ldots ,z(x_{k}))=\frac {\exp \left(-{\frac {1}{2}}({\mathbf {z} }-{\boldsymbol {\mu }})^{\mathrm {T} }{\boldsymbol {\Sigma }}(\mathbf {x})^{-1}({\mathbf {z} }-{\boldsymbol {\mu }})\right)}{\sqrt {(2\pi )^{k}|{\boldsymbol {\Sigma }}(\mathbf {x})|}}
\end{equation}
with $\mathbf {z}=(z(x_{1}),\ldots ,z(x_{k}))\in \mathbb{R}^{k}$, the mean ${\boldsymbol\mu = 0} \in \mathbb{R}^{k}$ in this study, and a squared-exponential correlation ${\boldsymbol {\Sigma }}(x_i,x_j) = \sigma^2 \exp( -|x_i - x_j|^2/a^2 )$, where the variance $\sigma^2$ is set to $1$, $a$ is the correlation range that is known and varied, and $x_i$ are spatial grid-points of the 2D field images. 
\begin{figure}
\centering
\includegraphics[scale=.33]{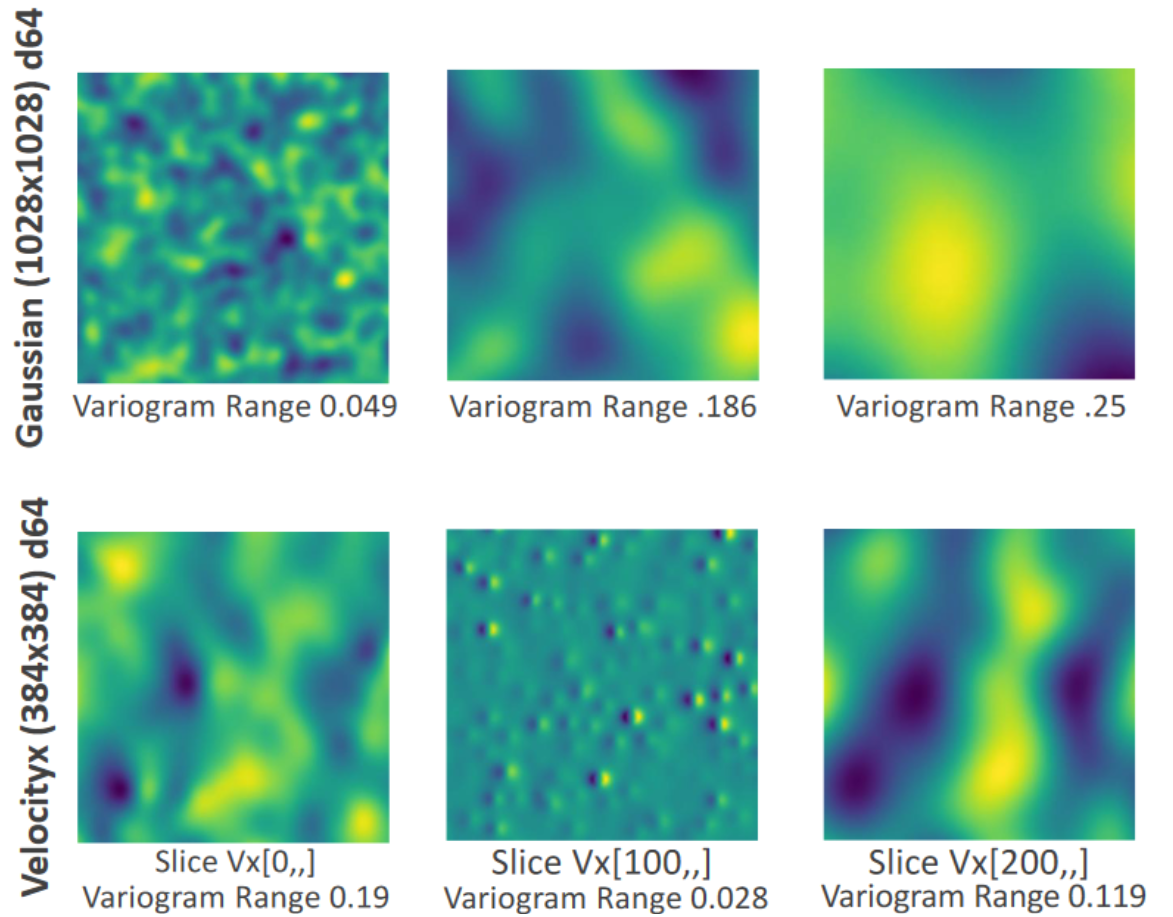}
\vspace{-10pt}
\caption{Original images of 2D Gaussian fields (top) and Miranda dataset 2D-slices (bottom)}
\label{original_slices}
\vspace{-10pt}
\end{figure}

In our evaluation, we consider two types of synthetic Gaussian datasets: single correlation range fields and multiple correlation range fields.
We consider single correlation range Gaussian fields as a proof of concept that gives us a high degree of control over correlation structures in the data.
However, because application datasets are likely to exhibit complex correlation structure as multiple correlation ranges, we consider also multiple range Gaussian fields.
For the multi-range correlation, we generate Gaussian fields with two distinct correlation ranges contributing equally to the total field.
This provides a controlled case with increased complexity. 

The final dataset is generated from the Miranda\cite{capello_zhao_di_tao_bessac_chen} code, designed for hydrodynamical large turbulence simulations. These data are more complex than the Gaussian fields due to multiple correlation ranges and complex dependencies. 
These original 3D data dimensions are 256$\times$384$\times$384. The 3D datasets are split into separate 2D slices based on equally spaced slices along the first dimension. In this paper, we use slices of the \textit{velocityx} variable as shown in Figure \ref{original_slices}.

\subsection{Compressors and Software}\label{sec:compressor}
Each lossy compressor in Table \ref{tab:software} is run with the following absolute error bounds: 1E-5, 1E-4, 1E-3, and 1E-2.
We chose the software and compressor versions from the latest available on Spack from a selection of leading error-bounded lossy compressors.
We use the absolute error bound because it is supported by each of the leading error-bounded lossy compressors.
Additionally, there are formal equivalences between the absolute error bound mode and other error bounds modes such as the value range relative error bound mode which are used by compressors such as SZ~\cite{Liang2018}.
Therefore, considering the compressablity given an absolute error bound is generalizable to other kinds of bounds.
All experiments are run on Clemson's Palmetto cluster using a node with two 32 core Intel(R) Xeon(R) Gold 6148 CPU @ 2.40GHz and 384 GB of RAM. The OS is Linux CentOS 8 with compiler GCC 8.4.1. 
Additional software and packages used in the study are listed in Table \ref{tab:software}. 
\begin{table}
\begin{adjustbox}{center}
\begin{tabular}{ ccc }
\textbf{Software} & \textbf{Version} & \textbf{Purpose} \\\toprule
SZ\cite{Liang2018} & @2.1.11.1 & lossy compressor \\ 
ZFP\cite{Lindstrom2006} & @0.5.5 & lossy compressor \\
MGARD\cite{Ainsworth2019} & @0.1.0 & lossy compressor \\
gstat\cite{Pebesma2004} & @2.0-7 & obtain variogram range \\
numpy\cite{harris2020array} & @1.21.1 & polyfit function to graph the curves \\
Libpressio\cite{underwood_di_calhoun_cappello_2020} & @0.70.0 & compress and measure the data \\\bottomrule
\end{tabular}
\end{adjustbox}
\caption{Compressors and software used for the study}
\label{tab:software}
\vspace{-20pt}
\end{table}

\subsection{Compression Statistics and Statistical Methods}\label{sec:stats}

\subsubsection{Compression Statistics}
Compression ratio, the ratio of the uncompressed data size by the compressed data size, is used to compare the different compressors and their efficiency. Compression ratio depends on: error bound, compressor used, and correlation structures within the data.  The compression ratio is comparable between different compressors and error bounds. 
In the following, this quantity is computed on the studied datasets from Section \ref{sec:data} for different compressors and error bounds, and investigated as function of a measure of several correlation statistics of data computed through the variogram range described below.

\subsubsection{Variogram study}
In the following section, we compute the empirical variogram of each 2D data-slice from the datasets described in Section \ref{sec:data} and based on the Euclidean distance between grid-points. % assigned to coordinates respecting their visual distance. 
The corresponding range $a$ is then estimated and reported in the following section as  estimated variogram range.  
In particular, we estimate the variogram ranges on the entire 2D field in order to assess the overall correlation structure of the fields. 
However, this is insufficient to characterize local heterogeneity in datasets, hence we compute the variogram ranges in windows of a given size that cover the entire 2D field in a tiled fashion \cite{bessac2019}. 
More specifically, we compute and report the standard deviation of  variogram ranges estimated over the windows covering the entire field. 
This statistic provides information on the spatial diversity and spread of local correlations present in the data.

\section{Experimental Results}\label{sec:compressability:results}

\subsection{Compressibility and global correlation}

\begin{figure}
\centering
{\includegraphics[scale=.27]{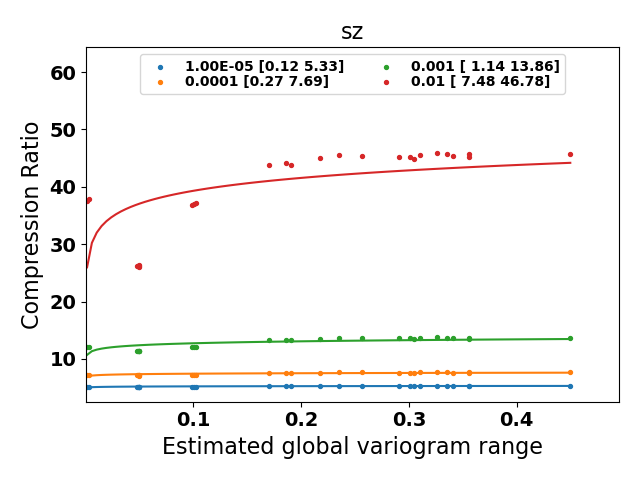}} \hspace{-10pt} \includegraphics[scale=.27]{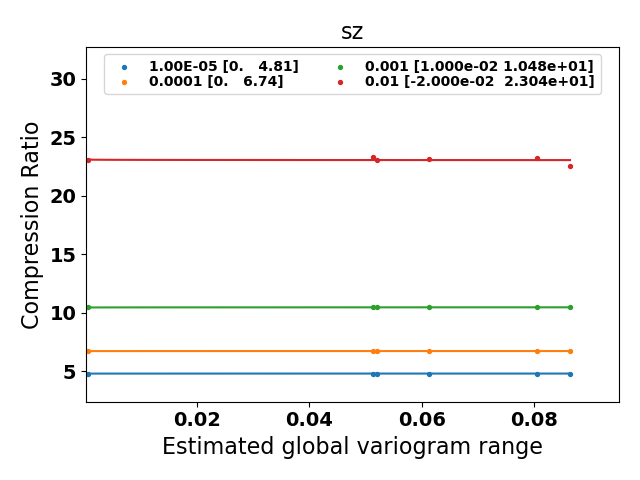} \\ 
{\includegraphics[scale=.27]{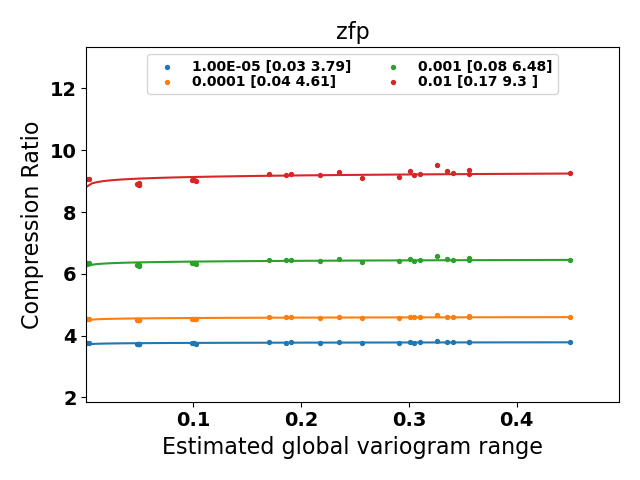}} \hspace{-10pt} 
\includegraphics[scale=.27]{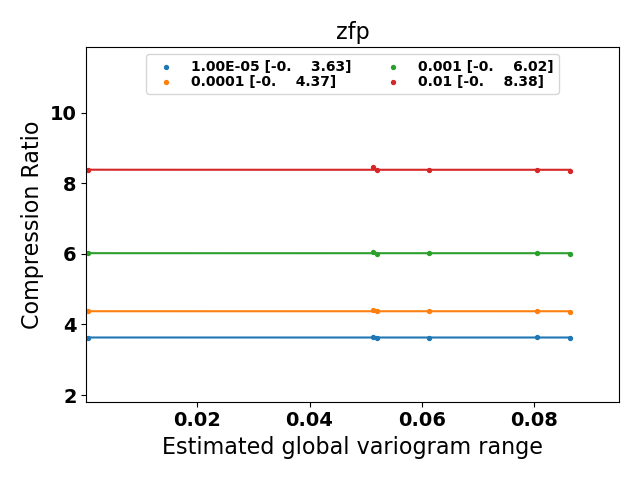} \\ 
{\includegraphics[scale=.27]{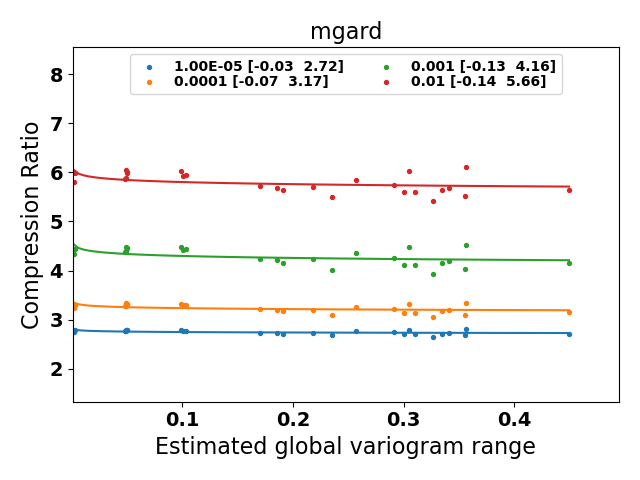}} \hspace{-10pt} \includegraphics[scale=.27]{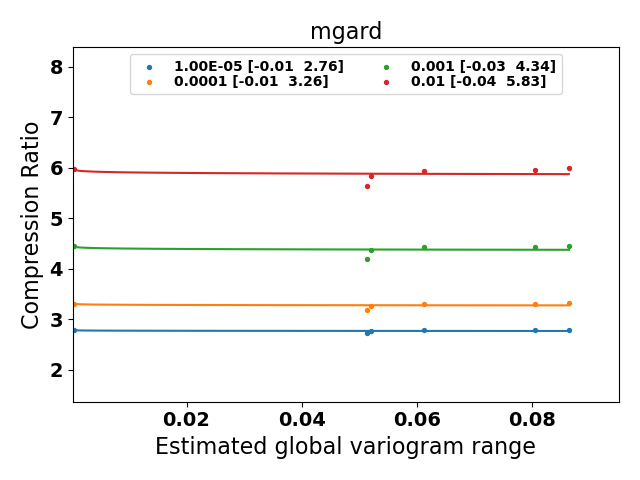} 
\vspace{-10pt}
\caption{Compression ratios against estimated variogram range for 2D Gaussian fields with single correlation range (left) and multi-range correlation (right). 
Each color is associated with  an error bound.
Empirical calculations are depicted with  dots and fitted logarithmic regressions are shown with solid lines. 
Estimated logarithmic regression coefficients $\alpha$ and $\beta$ are given in the legend.}
\label{results_gaussian}
\vspace{-10pt}
\end{figure}

In the following figures, the variogram range estimated on entire 2D fields is referred  to as ``Estimated global variogram range'', the standard deviation of locally estimated variogram range on $32\times 32$ windows is denoted as ``Std  estimated of local variogram range (H=32)''. 
Finally, an additional statistic are considered to illustrate future research directions. 
It consists in the standard deviation of locally extracted SVD thresholds on windows of size $32\times 32$. Thresholds correspond to the  number of required singular modes to recover $99\%$ of the variance of the initial field. 
This statistic is  referred to as ``Std of  truncation level of local SVD (H=32)''. 

\begin{figure}
\centering
{\includegraphics[scale=.27]{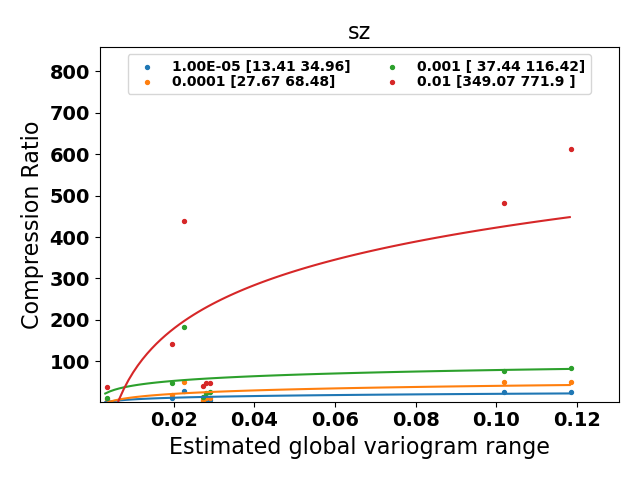}} \hspace{-10pt} 
 \includegraphics[scale=.27]{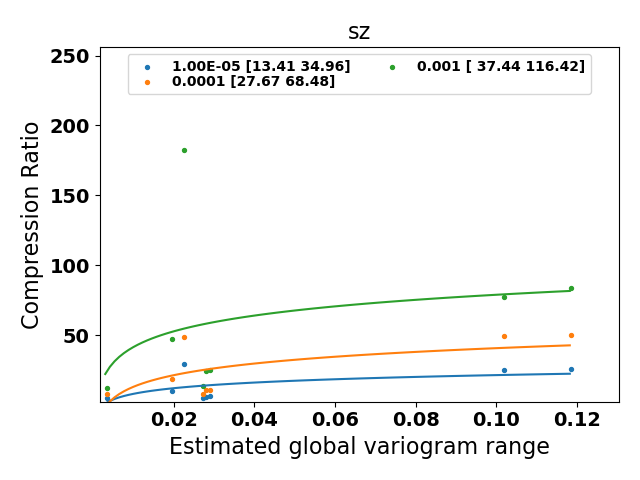}  \\
{\includegraphics[scale=.27]{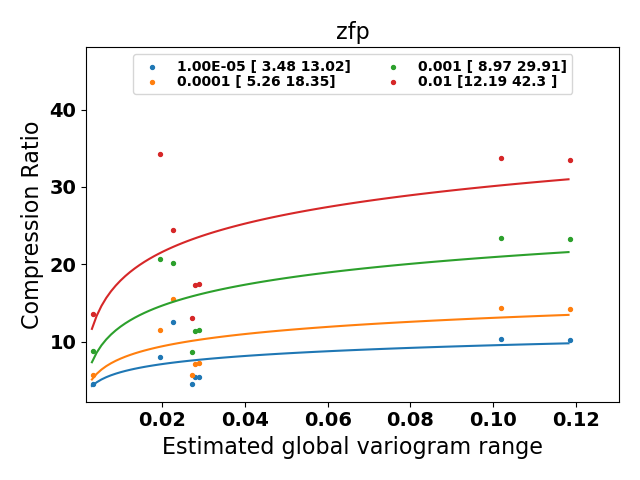}} \hspace{-10pt} 
{\includegraphics[scale=.27]{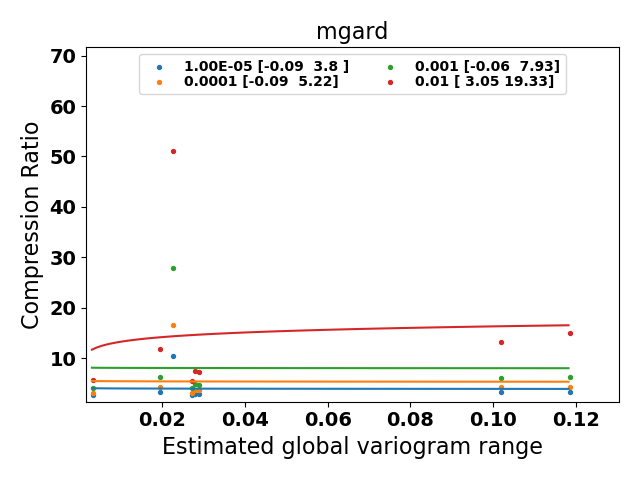}} 
\vspace{-10pt}
\caption{Compression ratios against estimated variogram range for Miranda \textit{velocityx}. Empirical calculations are depicted with  dots and fitted logarithmic regressions are shown with solid lines. 
Each color is associated with  an error bound. 
Estimated logarithmic regression coefficients $\alpha$ and $\beta$ are given in the legend. 
Due to the large spread of compression ratios for SZ (top left panel), the top right panel corresponds to results for error bounds strictly below 1E-2 in order to ease the reading.}
\label{results_velocityx}
\vspace{-10pt}
\end{figure}

In Figures \ref{results_gaussian} and \ref{results_velocityx} , the compression ratios for SZ\cite{Liang2018} and ZFP\cite{Lindstrom2006} (top panels) are plotted against the variogram range estimated  on  entire 2D fields.   
These statistics are respectively estimated on the synthetic Gaussian fields (left: single-range correlation and right: multi-range correlation) and Miranda datasets. 
Multiple curves shown on each panel correspond to different error bounds of the lossy compressor. 
As the estimated global variogram range increases indicating stronger dependence between spatial points, dataset variability decreases yielding smoother, more compressible data.  
This increasing relationship between compression ratio and variogram range exhibits a plateau for highly correlated data  (large variogram ranges) suggesting a limit in compressibility of the data for a given error bound and compressor. 
Note that this trend is less visible on the  multi-range correlation Gaussian fields (right column of Figure \ref{results_gaussian}) due to the equal contributions of each correlation to the total field, preventing  any  dominant correlation pattern to  prevail and thus to be characterized  properly by  the variogram range of the entire field. 
Compression ratios for MGARD\cite{Ainsworth2019} are less sensitive to the dependencies  to correlation ranges present in the data which is likely due to the global scope of the compressor. 

Finally, in order to quantify these relationships and compare them across different compressors and error bounds, the compression ratios are fitted by least-squares to logarithmic regressions of the estimated variogram range $a$: $CR = \alpha +\beta \log(a) +\epsilon$, where $CR$ is the compression ratio, $a$ the estimated variogram range, and $\alpha$ and $\beta$ are estimated coefficients and reported in the legend box of each panel.   
The fitted logarithmic regressions show a good match  to the datapoints indicating in most cases a logarithmic dependence of the compression ratios to the estimated variogram ranges. 
Lower compression error bounds exhibit lower variance of datapoints around the fitted curve and fewer outliers. 
Regressions fit better the single-scale correlation Gaussian fields (Figure \ref{results_gaussian} left column) than multi-range correlation Gaussian fields and the Miranda data due to their lower complexity that is captured reasonably by the global variogram range.  
In particular, the fitting on datapoints from  Miranda data show more dispersion around the fitted curves but a matching trend. 

The variogram range estimated on each entire 2D field represents an average correlation range observed on the entire field and is not suited to characterize spatial local  heterogeneity nor multiple scales that can be present in complex scientific datasets, such as in Miranda datasets or multi-range Gaussian fields. 
In the following section, we explore local correlations and  their link to compression ratios. 
%Future efforts will be target these characterizations. 

\subsection{Compressibility and local correlation}
\begin{figure}
\centering
\includegraphics[scale=.27]{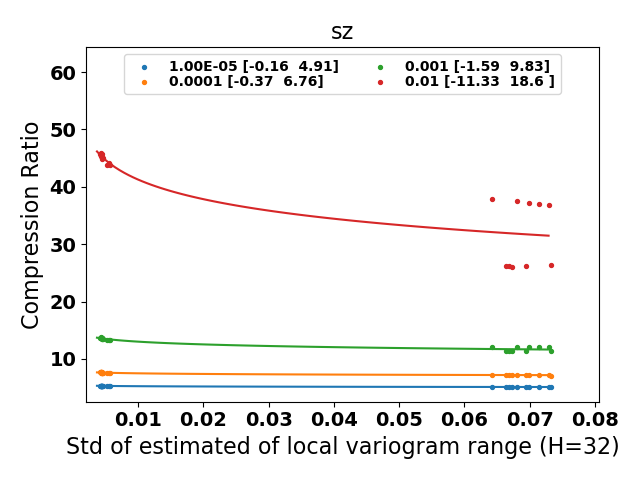} \hspace{-10pt} 
\includegraphics[scale=.27]{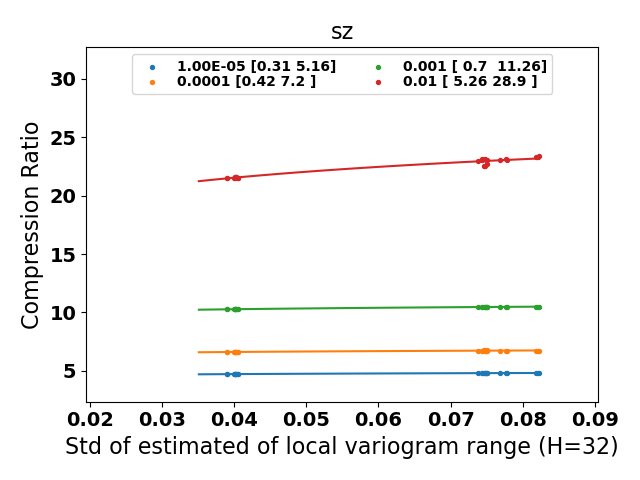} \\ 
\includegraphics[scale=.27]{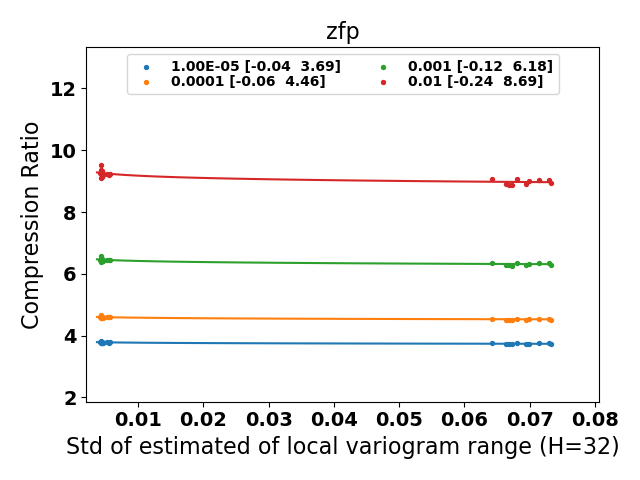} \hspace{-10pt} 
\includegraphics[scale=.27]{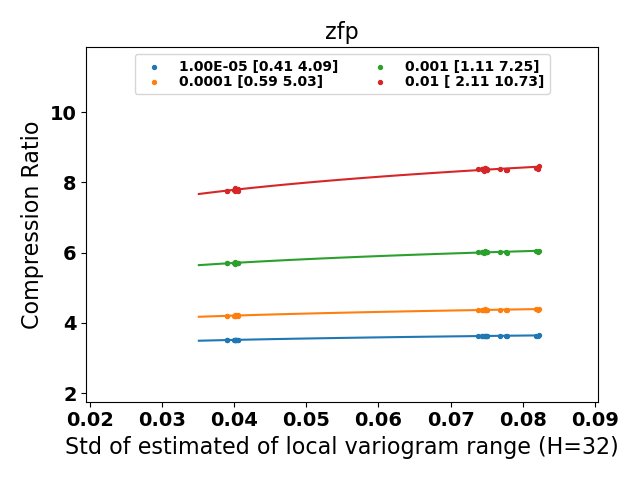} \\ 
\includegraphics[scale=.27]{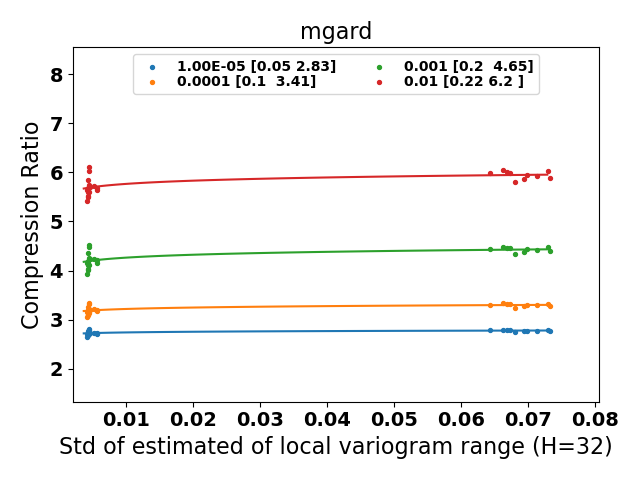} \hspace{-10pt} 
\includegraphics[scale=.27]{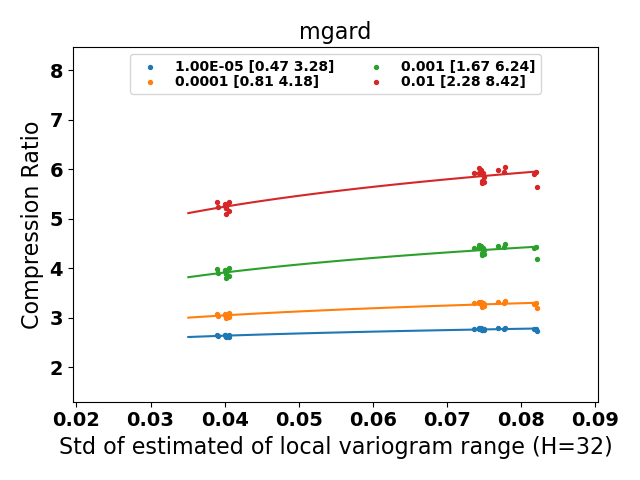}
\vspace{-10pt}
\caption{Compression ratios against standard deviation of the local variogram range  for single range  correlation Gaussian fields (right) and for multi-range correlation Gaussian fields (right). 
Initial step towards the use of decomposition techniques to analysis fields with multiscale patterns. 
Empirical calculations are depicted with dots and fitted logarithmic regressions are shown with solid lines. Each color is associated with an error bound. 
Estimated logarithmic regression coefficients $\alpha$ and $\beta$ are given in the legend. }
\vspace{-15pt}
\label{fig:results_localvariog_gaussian}
\end{figure}

As the variogram ranges estimated on the entire 2D fields may not provide enough insights  on local heterogeneity, we estimate the variogram ranges on local windows ($32 \times 32$) tiling each entire 2D-field. 
Additionally,  exploring local heterogeneity is important since many leading error-bounded lossy compressors exploit some notions of locality in their algorithms as discussed in
Section~\ref{sec:compressablity:compressors}. 
In Figure \ref{results_gaussian}, as the global variogram range increased, the compression ratio fails to have a slope of greater than absolute value of 1E-2 for mutli-range Gaussian fields (right column).  
This illustrates the shortcomings of the global variogram range statistic while dealing with multi-range correlation datasets. 
Figure \ref{fig:results_localvariog_gaussian} shows the compression ratios computed on single-range correlation Gaussian fields (left) and multi-range  correlation fields (right) as  a function  of the local variogram ranges and the left column of Figure  \ref{fig:results_localstats_miranda} shows the statistics computed on the Miranda fields. 
Both  figures corroborate and illustrate the benefit of considering local information for complex datasets to be better explain compression ratios by local variograms rather than global ones. 
In particular,  we observe that Figures \ref{results_gaussian} and \ref{results_velocityx} depict several close values of compression ratios for close variogram  ranges, whereas this effect is  less visible in  Figures \ref{fig:results_localvariog_gaussian} and \ref{fig:results_localstats_miranda} indicating a stronger exploratory skill of the local statistic to the compression ratios. 
However, results for  the  single-range correlation Gaussian fields show  a weaker sensitivity of  the  compression ratios  to this local  statistic. 
This might suggest the need to use several  statistics to provide appropriate explanatory skills. 
Future work will  explore this  path. 

\subsection{Towards multiscale analysis and summary statistics}

\begin{figure}
\centering
 \includegraphics[scale=.27]{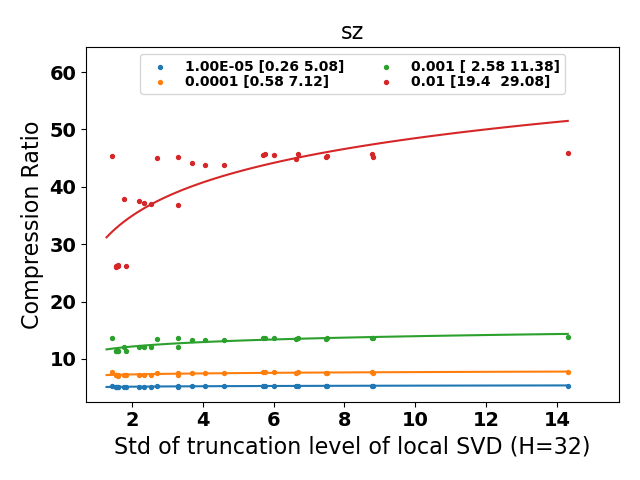} \hspace{-8pt} 
 \includegraphics[scale=.27]{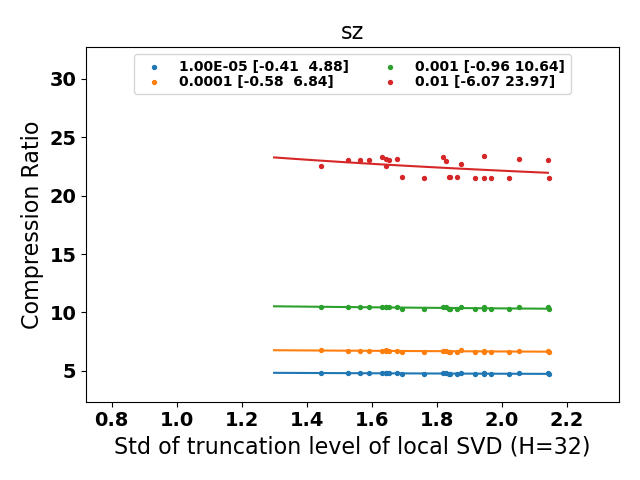} 
 \includegraphics[scale=.27]{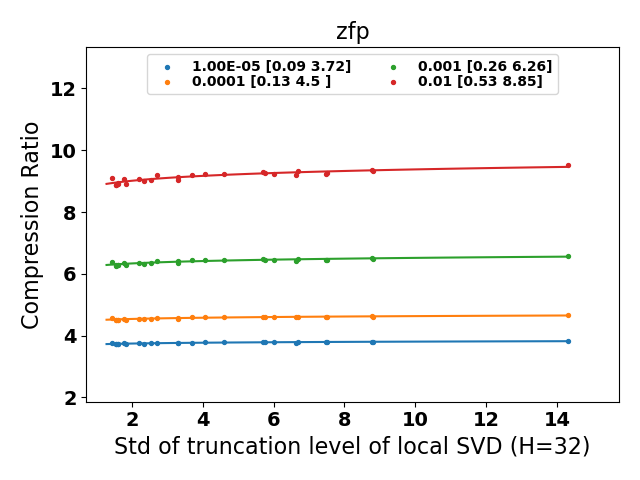} \hspace{-8pt} 
 \includegraphics[scale=.27]{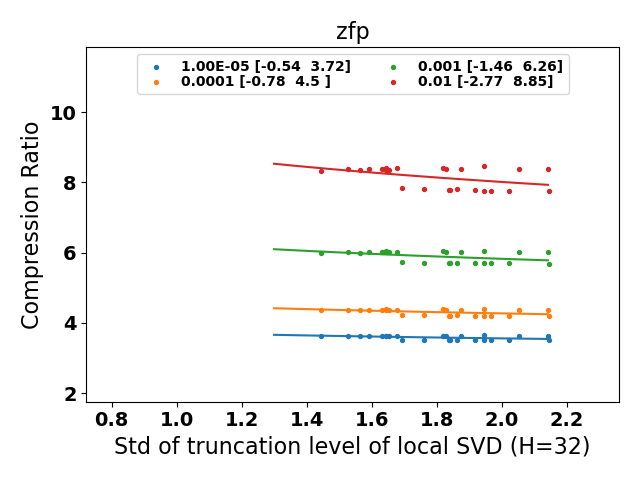} 
 \vspace{-10pt}
\caption{Compression ratios against standard deviation of the truncation level of local SVD for  single  scale  Gaussian fields (left) and for multiscale Gaussian datasets (right). 
Empirical calculations are depicted with dots and fitted logarithmic regressions are shown with solid lines. 
Each color is associated with an error bound.
Estimated logarithmic regression coefficients $\alpha$ and $\beta$ are given in the legend. 
}
\label{fig:results_localsvd_gaussian}
\vspace{-15pt}
\end{figure}

\begin{figure}
\centering
\includegraphics[scale=.27]{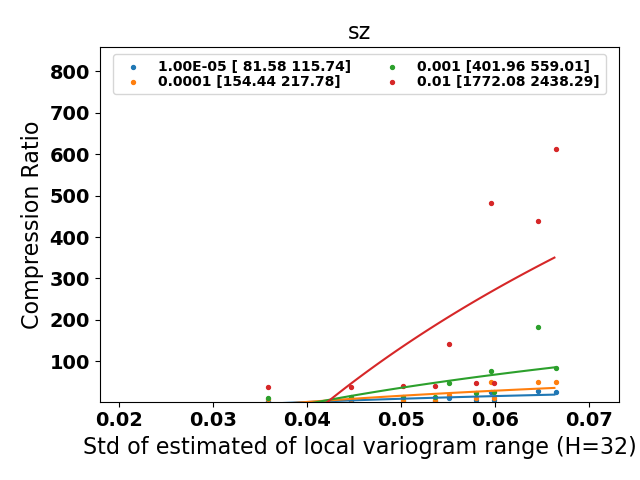} \hspace{-8pt} 
\includegraphics[scale=.27]{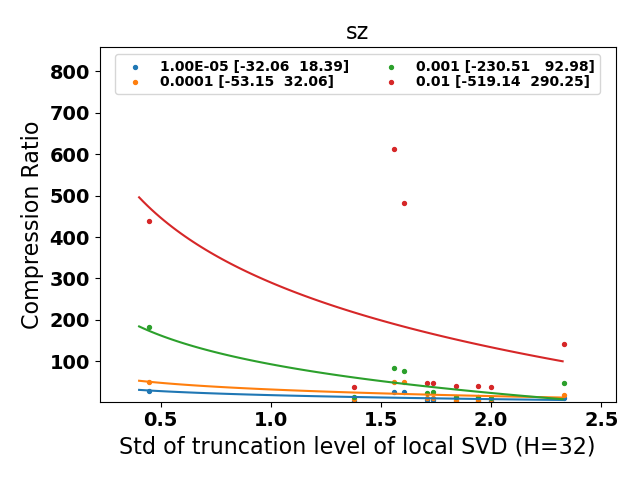} \\
\vspace{-10pt}
\includegraphics[scale=.27]{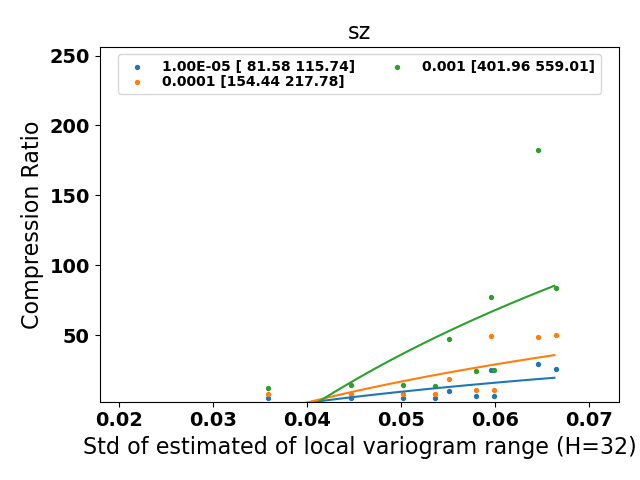} \hspace{-8pt} 
\includegraphics[scale=.27]{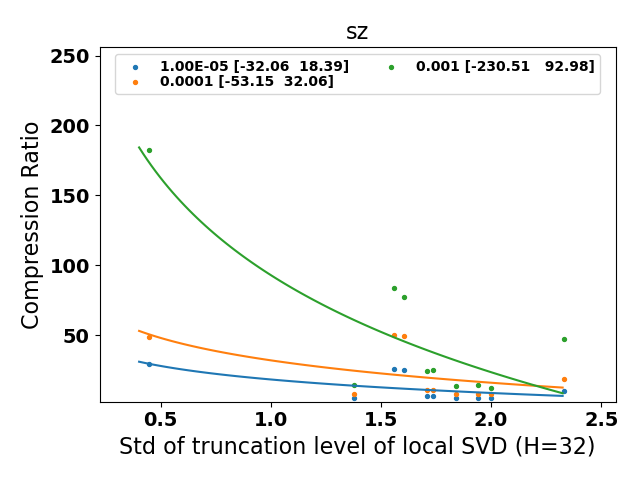} \\
\vspace{-10pt}
\includegraphics[scale=.27]{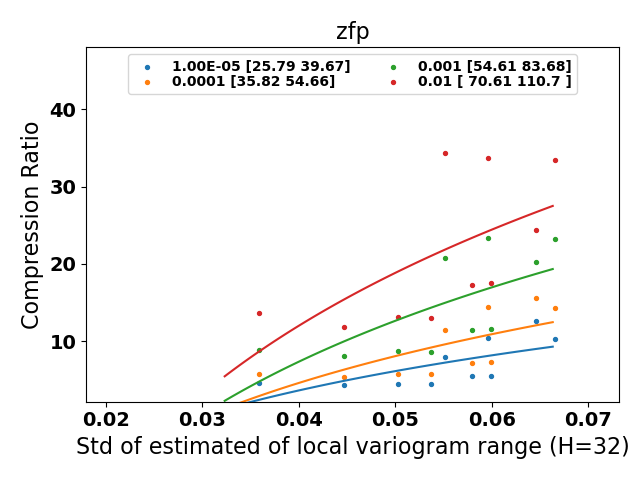} \hspace{-8pt} 
\includegraphics[scale=.27]{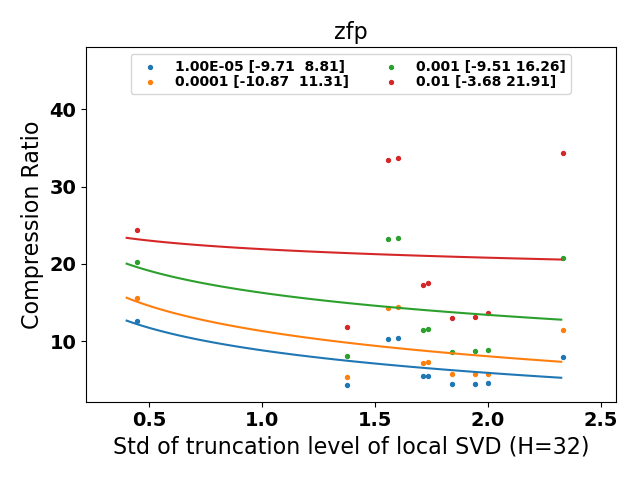} \\
\vspace{-10pt}
\caption{Compression ratios against local statistics, standard deviation of the local variogram range (left) and standard deviation of the truncation level of local SVD (left) for Miranda \textit{velocityx} 2D field. 
Empirical calculations are depicted with dots and fitted logarithmic regressions are shown with solid lines. Each color is associated with an error bound. 
Estimated logarithmic regression coefficients $\alpha$ and $\beta$ are given in the legend. 
Due to the large spread of compression ratios for SZ, the central panels correspond to results for error bounds strictly below 1E-2 in order to ease the reading. }
\label{fig:results_localstats_miranda}
\vspace{-15pt}
\end{figure}

Figure \ref{fig:results_localsvd_gaussian} and the right column of Figure \ref{fig:results_localstats_miranda} illustrate an initial step towards the use of decomposition techniques to analysis fields with multiscale patterns. In this  section, local SVD  are performed on the fields and summarized  via the standard deviation of locally required number of singular modes to capture $99\%$ of  the variance  of each  local window. 
This  local statistic, larger values of required singular modes are associated with less compressible fields  so we expect mostly decreasing relationship of compression  ratios to this local statistic. 
For simplicity, as this  section provides insights to future  works, we omit  the  compressor MGARD  as  it showed less sensitivity to the previous statistics. 
Figure \ref{fig:results_localsvd_gaussian} and the right column of Figure \ref{fig:results_localstats_miranda} indicate that this  local statistic provide  a  more diverse representation of the data as seen by the  span of unique values  over  the x-axis, than  the two other statistics based on variogram.  
This statistic tend to exhibit several  relating trends to compression ratio, highlighting a need  to refine this statistic.   
Future work will explore transformation of  this  statistic or other representations based local SVDs and variograms, as both methods provided various information to explain compression ratios.

\section{Conclusions and Future Work}

Establishing the limit for lossy compression would allow for the maximum efficiency for compressing and storing large scientific datasets. 
Our work represent a first step toward this goal.  
We have demonstrated that estimated global and local variogram ranges can explain compression ratio in a logarithmic fashion for some compressors and given error bounds.
This hypothesis was illustrated on synthetic Gaussian fields providing a proof of concept and corroborated by results on a user scientific dataset from Miranda. 
With the studied datasets, SZ and ZFP seem to utilize the global and local spatial correlation ranges in a logarithmic fashion with various coefficients showing the strength of the dependence. MGARD seems less sensitive to these trends in its compression ratios. 

Since heterogeneous (non-stationary) and multiscale correlations in the data may be mis-represented by the global spatial variogram, other statistics, including local variograms and local SVD, have been studied to address these issues. 
In continuation of this research, there are a few goals for the future: i) explore more complex dependent variables (local correlation combined with multiscale statistics based on decomposition) as candidate predictors,  ii) create  more complex synthetic multiscale 2D Gaussian fields and iii) create a model of compression ratio based on correlation metrics and error bound. 
Future work will investigate the effects of correlation structures  on quality metrics of reconstructed data such as PSNR along with  a design of the statistics to a 3D context.  
% As a final note, building scalar statistics that are representative of spatial or multidimensional fields is a current challenge in many applications \cite{buschow2021}. 

Another aspect for future work is how to quickly assess this metric.
The currently implementation relies on the singular value decomposition which is slow relative to modern compressors.
We plan to leverage a sampling approach similar to prior work\cite{taoOptimizingLossyCompression2019,lu2018understanding}. We are hopeful that increasing levels of sampling by block can provide an increasingly accurate proxy for our metric.

\section*{Acknowledgments}
Clemson University is acknowledged for generous allotment of compute time on the Palmetto cluster. This material is based upon work supported by the National Science Foundation under Grant No. SHF-1910197.

% use bibtex to make reference formatting easy!
\bibliographystyle{IEEEtran}
\bibliography{refs.bib}

\begin{thebibliography}{10}
\providecommand{\url}[1]{#1}
\csname url@samestyle\endcsname
\providecommand{\newblock}{\relax}
\providecommand{\bibinfo}[2]{#2}
\providecommand{\BIBentrySTDinterwordspacing}{\spaceskip=0pt\relax}
\providecommand{\BIBentryALTinterwordstretchfactor}{4}
\providecommand{\BIBentryALTinterwordspacing}{\spaceskip=\fontdimen2\font plus
\BIBentryALTinterwordstretchfactor\fontdimen3\font minus
  \fontdimen4\font\relax}
\providecommand{\BIBforeignlanguage}[2]{{%
\expandafter\ifx\csname l@#1\endcsname\relax
\typeout{** WARNING: IEEEtran.bst: No hyphenation pattern has been}%
\typeout{** loaded for the language `#1'. Using the pattern for}%
\typeout{** the default language instead.}%
\else
\language=\csname l@#1\endcsname
\fi
#2}}
\providecommand{\BIBdecl}{\relax}
\BIBdecl

\bibitem{Cappello2019}
\BIBentryALTinterwordspacing
F.~Cappello, S.~Di, S.~Li, X.~Liang, A.~M. Gok, D.~Tao, C.~H. Yoon, X.-C. Wu,
  Y.~Alexeev, and F.~T. Chong, ``Use cases of lossy compression for
  floating-point data in scientific data sets,'' \emph{The International
  Journal of High Performance Computing Applications}, vol.~33, no.~6, pp.
  1201--1220, Jul. 2019. [Online]. Available:
  \url{https://doi.org/10.1177/1094342019853336}
\BIBentrySTDinterwordspacing

\bibitem{Shannon1948}
\BIBentryALTinterwordspacing
C.~E. Shannon, ``A mathematical theory of communication,'' \emph{Bell System
  Technical Journal}, vol.~27, no.~3, pp. 379--423, Jul. 1948. [Online].
  Available: \url{https://doi.org/10.1002/j.1538-7305.1948.tb01338.x}
\BIBentrySTDinterwordspacing

\bibitem{Liang2018}
\BIBentryALTinterwordspacing
X.~Liang, S.~Di, D.~Tao, S.~Li, S.~Li, H.~Guo, Z.~Chen, and F.~Cappello,
  ``Error-controlled lossy compression optimized for high compression ratios of
  scientific datasets,'' in \emph{2018 {IEEE} International Conference on Big
  Data (Big Data)}.\hskip 1em plus 0.5em minus 0.4em\relax {IEEE}, Dec. 2018.
  [Online]. Available: \url{https://doi.org/10.1109/bigdata.2018.8622520}
\BIBentrySTDinterwordspacing

\bibitem{Lindstrom2006}
\BIBentryALTinterwordspacing
P.~Lindstrom and M.~Isenburg, ``Fast and efficient compression of
  floating-point data,'' \emph{{IEEE} Transactions on Visualization and
  Computer Graphics}, vol.~12, no.~5, pp. 1245--1250, Sep. 2006. [Online].
  Available: \url{https://doi.org/10.1109/tvcg.2006.143}
\BIBentrySTDinterwordspacing

\bibitem{capello_zhao_di_tao_bessac_chen}
\BIBentryALTinterwordspacing
F.~Capello, K.~Zhao, S.~Di, D.~Tao, J.~Bessac, and Z.~Chen, Jun 2018. [Online].
  Available: \url{https://sdrbench.github.io/}
\BIBentrySTDinterwordspacing

\bibitem{Ainsworth2019}
\BIBentryALTinterwordspacing
M.~Ainsworth, O.~Tugluk, B.~Whitney, and S.~Klasky, ``Multilevel techniques for
  compression and reduction of scientific data---the multivariate case,''
  \emph{{SIAM} Journal on Scientific Computing}, vol.~41, no.~2, pp.
  A1278--A1303, Jan. 2019. [Online]. Available:
  \url{https://doi.org/10.1137/18m1166651}
\BIBentrySTDinterwordspacing

\bibitem{matheron1963}
G.~Matheron, ``Principles of geostatistics,'' \emph{Economic geology}, vol.~58,
  no.~8, pp. 1246--1266, 1963.

\bibitem{gelfand2010}
A.~E. Gelfand, P.~Diggle, P.~Guttorp, and M.~Fuentes, \emph{Handbook of spatial
  statistics}.\hskip 1em plus 0.5em minus 0.4em\relax CRC press, 2010.

\bibitem{abry1998}
P.~Abry, D.~Veitch, and P.~Flandrin, ``Long-range dependence: Revisiting
  aggregation with wavelets,'' \emph{Journal of Time Series Analysis}, vol.~19,
  no.~3, pp. 253--266, 1998.

\bibitem{cabrieto2017}
J.~Cabrieto, F.~Tuerlinckx, P.~Kuppens, M.~Grassmann, and E.~Ceulemans,
  ``Detecting correlation changes in multivariate time series: A comparison of
  four non-parametric change point detection methods,'' \emph{Behavior research
  methods}, vol.~49, no.~3, pp. 988--1005, 2017.

\bibitem{bessac2016}
J.~Bessac, P.~Ailliot, J.~Cattiaux, and V.~Monbet, ``Comparison of hidden and
  observed regime-switching autoregressive models for (u,v)-components of wind
  fields in the {N}ortheast {A}tlantic,'' \emph{Advances in Statistical
  Climatology, Meteorology and Oceanography}, vol.~2, no.~1, pp. 1--16, 2016.

\bibitem{addison2017}
P.~S. Addison, \emph{The illustrated wavelet transform handbook: introductory
  theory and applications in science, engineering, medicine and finance}.\hskip
  1em plus 0.5em minus 0.4em\relax CRC press, 2017.

\bibitem{hannachi2007}
A.~Hannachi, I.~T. Jolliffe, and D.~B. Stephenson, ``Empirical orthogonal
  functions and related techniques in atmospheric science: A review,''
  \emph{International Journal of Climatology: A Journal of the Royal
  Meteorological Society}, vol.~27, no.~9, pp. 1119--1152, 2007.

\bibitem{gavish2014}
M.~Gavish and D.~L. Donoho, ``The optimal hard threshold for singular values is
  $4\backslash \sqrt{3}$,'' pp. 5040 -- 5053, 2014.

\bibitem{lu2018understanding}
T.~Lu, Q.~Liu, X.~He, H.~Luo, E.~Suchyta, J.~Choi, N.~Podhorszki, S.~Klasky,
  M.~Wolf, T.~Liu \emph{et~al.}, ``Understanding and modeling lossy compression
  schemes on hpc scientific data,'' in \emph{2018 IEEE International Parallel
  and Distributed Processing Symposium (IPDPS)}.\hskip 1em plus 0.5em minus
  0.4em\relax IEEE, 2018, pp. 348--357.

\bibitem{qin2020estimating}
Z.~Qin, J.~Wang, Q.~Liu, J.~Chen, D.~Pugmire, N.~Podhorszki, and S.~Klasky,
  ``Estimating lossy compressibility of scientific data using deep neural
  networks,'' \emph{IEEE Letters of the Computer Society}, vol.~3, no.~1, pp.
  5--8, 2020.

\bibitem{hanDataMiningConcepts2011}
J.~Han, J.~Pei, and M.~Kamber, \emph{Data Mining: Concepts and
  Techniques}.\hskip 1em plus 0.5em minus 0.4em\relax {Elsevier}, 2011.

\bibitem{taoOptimizingLossyCompression2019}
D.~Tao, S.~Di, X.~Liang, Z.~Chen, and F.~Cappello, ``Optimizing {{Lossy
  Compression Rate}}-{{Distortion}} from {{Automatic Online Selection}} between
  {{SZ}} and {{ZFP}},'' \emph{IEEE Transactions on Parallel and Distributed
  Systems}, vol.~30, no.~8, pp. 1857--1871, Aug. 2019.

\bibitem{lindstrom2014}
P.~Lindstrom, ``Fixed-rate compressed floating-point arrays,'' \emph{IEEE
  Transactions on Visualization and Computer Graphics}, vol.~20, no.~12, pp.
  2674--2683, 2014.

\bibitem{moon2017}
A.~Moon, J.~Kim, J.~Zhang, and S.~W. Son, ``Lossy compression on iot big data
  by exploiting spatiotemporal correlation,'' in \emph{2017 IEEE High
  Performance Extreme Computing Conference (HPEC)}.\hskip 1em plus 0.5em minus
  0.4em\relax IEEE, 2017, pp. 1--7.

\bibitem{yu2021}
Z.~Yu, D.~Li, Z.~Zhang, W.~Luo, Y.~Liu, Z.~Wang, and L.~Yuan, ``Lossy
  compression of earth system model data based on a hierarchical tensor with
  adaptive-hgfdr (v1. 0),'' \emph{Geoscientific Model Development}, vol.~14,
  no.~2, pp. 875--887, 2021.

\bibitem{Pebesma2004}
\BIBentryALTinterwordspacing
E.~J. Pebesma, ``Multivariable geostatistics in s: the gstat package,''
  \emph{Computers {\&} Geosciences}, vol.~30, no.~7, pp. 683--691, Aug. 2004.
  [Online]. Available: \url{https://doi.org/10.1016/j.cageo.2004.03.012}
\BIBentrySTDinterwordspacing

\bibitem{harris2020array}
\BIBentryALTinterwordspacing
C.~R. Harris, K.~J. Millman, S.~J. van~der Walt, R.~Gommers, P.~Virtanen,
  D.~Cournapeau, E.~Wieser, J.~Taylor, S.~Berg, N.~J. Smith, R.~Kern, M.~Picus,
  S.~Hoyer, M.~H. van Kerkwijk, M.~Brett, A.~Haldane, J.~F. del R{\'{i}}o,
  M.~Wiebe, P.~Peterson, P.~G{\'{e}}rard-Marchant, K.~Sheppard, T.~Reddy,
  W.~Weckesser, H.~Abbasi, C.~Gohlke, and T.~E. Oliphant, ``Array programming
  with {NumPy},'' \emph{Nature}, vol. 585, no. 7825, pp. 357--362, Sep. 2020.
  [Online]. Available: \url{https://doi.org/10.1038/s41586-020-2649-2}
\BIBentrySTDinterwordspacing

\bibitem{underwood_di_calhoun_cappello_2020}
R.~Underwood, S.~Di, J.~C. Calhoun, and F.~Cappello, ``Fraz: A generic
  high-fidelity fixed-ratio lossy compression framework for scientific
  floating-point data,'' \emph{2020 IEEE International Parallel and Distributed
  Processing Symposium (IPDPS)}, 2020.

\bibitem{bessac2019}
\BIBentryALTinterwordspacing
J.~Bessac, A.~H. Monahan, H.~M. Christensen, and N.~Weitzel, ``Stochastic
  parameterization of subgrid-scale velocity enhancement of sea surface
  fluxes,'' \emph{Monthly Weather Review}, vol. 147, no.~5, pp. 1447--1469,
  2019. [Online]. Available: \url{https://doi.org/10.1175/MWR-D-18-0384.1}
\BIBentrySTDinterwordspacing

\end{thebibliography}

\end{document}